\documentclass[preprint2]{aastex}
\shorttitle{HD~137510: an oasis in the brown dwarf desert.}
\shortauthors{Endl M., Hatzes A.P., Cochran W.D., et al.}

\begin{document}

\title{HD~137510: AN OASIS IN THE BROWN DWARF DESERT
\footnote{Based on observations made at McDonald Observatory and the 
Th\"uringer Landessternwarte Tautenburg}}
\author{Michael Endl} 
\affil{McDonald Observatory, The University of Texas at Austin, Austin, TX 78712, USA}
\email{mike@astro.as.utexas.edu}
\author{Artie P. Hatzes}
\affil{Th\"uringer Landessternwarte, D - 07778 Tautenburg, Germany}
\email{artie@tls-tautenburg.de}
\author{William D. Cochran, Barbara McArthur, Carlos Allende Prieto, Diane B. Paulson
\footnote{current address: Department of Astronomy, University of Michigan, 830 Dennison Ann Arbor MI 48109, USA}}
\affil{McDonald Observatory, The University of Texas at Austin, Austin, TX 78712, USA}
\email{wdc@astro.as.utexas.edu; mca@astro.as.utexas.edu; callende@astro.as.utexas.edu, apodis@umich.edu}
\and
\author{Eike Guenther, Ana Bedalov
\footnote{current address: Astrophysikalisches Institut, Universit\"at Jena, Schillerg\"asschen 2-3, D-07745 Jena, Germany}
}
\affil{Th\"uringer Landessternwarte, D - 07778 Tautenburg, Germany}
\email{guenther@tls-tautenburg.de; abedalov@tls-tautenburg.de}

\begin{abstract}

Since the beginning of precise Doppler surveys, which have had stunning success in detecting extrasolar
planetary companions, one surprising enigma has emerged: the relative paucity of spectroscopic 
binaries where the
secondary mass lies in between the stellar and planetary mass regime. This gap in the
mass function for close-in ($a<3-4$~AU) companions to solar-type stars is generally referred to as
the ``Brown Dwarf Desert''. Here we report the detection of a companion to HD~137510 (G0IV), with a 
minimum mass of $26~{\rm M}_{\rm Jupiter}$, moving in an eccentric orbit ($e=0.4$) with a 
period of $798$ days and an orbital semimajor axis of $1.85$~AU.
The detection is based on precise differential radial velocity (RV) data obtained by the McDonald
Observatory and Th\"uringer Landessternwarte Tautenburg planet search programs.    

\end{abstract}


\section{Introduction}

It was the common expectation that, with the increase in the precision of RV measurements, the
detected secondary masses of spectroscopic binaries will gradually decrease, first into the
brown dwarf mass range (${\rm M}<80~{\rm M}_{\rm Jup}$) and eventually into the planetary mass 
regime of roughly ${\rm M}<13~{\rm M}_{\rm Jup}$ (below the deuterium burning limit). 
Quite the opposite turned out to happen. 
The pioneering long-term Doppler survey of Campbell, Walker, \& Yang (1988), and later also in the southern 
hemisphere by Murdoch, Hearnshaw, \& Clark~(1993), did not reveal any brown dwarf 
companions to nearby solar-type stars, despite their sufficient sensitivity. 
There appeared to be an abrupt drop in the frequency of companions at the hydrogen burning
limit of $\approx 0.075~{\rm M}_{\odot}$.
The first very low-mass companion ($m \sin i = 11~{\rm M}_{\rm Jup}$)
detected by the RV technique is the companion to HD~114762 (Latham et al.~1989). 
The discoverers classified the object as a brown dwarf companion, but according to 
Cochran, Hatzes, \& Hancock~(1991) 
this companion might be in fact a late M dwarf due to a possible low value of the
inclination angle $i$. In the years after the discovery of the first unambiguous extrasolar
planet orbiting a main-sequence star by Mayor \& Queloz (1995), a total of more than a $100$ extrasolar 
planets have been accumulated by several precise Doppler searches all around the globe.   
Still, with a few exceptions (HD~127506: Mayor, Queloz, \& Udry~1998; HD~29587 \& HD~140913: 
Mazeh, Latham, \& Stefanik~1996; 
HD~168443 (second companion): Marcy \& Butler~2001; HD~184860: Vogt et al.~2002), 
the part of the mass function for spectroscopic
secondaries, in between the minimum hydrogen burning mass and the  
deuterium burning limit of ${\rm M}\approx13~{\rm M}_{\rm Jup}$, 
remained surprisingly sparsely populated. A notable exception is HD~10697, where Vogt et al.~(2000)
detected a companion with $m \sin i = 6.35~{\rm M}_{\rm Jup}$, while Zucker \& Mazeh (2000) determined
a true mass for the secondary of $40~{\rm M}_{\rm Jup}$ after combining the spectroscopic solution with
$Hipparcos$ astrometry. Udry et al.~(2002) found 4 companions with minimum masses close to the
planet/brown dwarf border, which could be either ``superplanets'' or real brown dwarfs.
However, most of the candidate brown dwarf secondaries considered
in the Halbwachs et al.~(2000) study turned out to be stellar mass companions, again after combination 
with $Hipparcos$ astrometry. 

This suggests the existence of two distinctive binary formation processes for certain mass-ratios. 
The gap in the mass function between the stellar and planetary mass range for close binaries 
is generally dubbed the ``Brown Dwarf Desert'' (this desert does not appear to exist at wide 
separations, e.g. Gizis et al. 2001; Neuh\"auser \& Guenther 2004). 
The gap is particularily obvious as a sudden drop at the deuterium burning limit in the
substellar mass function for close companions. While planets are relatively common
as companions to solar-type stars, brown dwarfs are rare.  
This fact is especially surprising since brown dwarf companions are much easier
to detect, even for lower precision surveys, due to the larger RV amplitudes.
Moreover, these objects cover a more than 4 times larger mass range
than planetary companions, still their detected numbers per mass bin is extremely
low compared to lower and higher mass companions.

In this work we present the discovery of another possible ``oasis'' in this desert, a brown dwarf 
orbiting the G-type star HD~137510.   

\section{Stellar characteristics of HD~137510}

HD~137510 (= HR~5740 = HIP~75535) is a bright ($V=6.3$) G-type star at a distance of
$41.75\pm1.7$ pc, according to the $Hipparcos$ parallax of $23.95\pm0.94$ mas (ESA~1997).
In order to derive the stellar parameters for the primary we
used the (B-V)-${\rm T}_{\rm eff}$ calibration of Alonso, Arribas, \& Mart\'{\i}nez-Roger~(1996), 
compared $M_V$ and ${\rm T}_{\rm eff}$ with stellar evolution
calculations (e.g. Reddy et al.~2003), and performed a model atmosphere analysis of a 
high resolution, high signal-to-noise spectrum of HD~137510.
We find that the star has a mass of ${\rm M} = 1.3\pm0.1~{\rm M}_{\odot}$, an effective temperature
${\rm T}_{\rm eff} = 5896 \pm 57$~K, a ${\rm log~g}$ of $4.0 \pm 0.1$,
a radius ${\rm R}$ of $1.9 \pm 0.2~{\rm R}_{\odot}$ and an age of $3.4 \pm 0.8$~Gyrs.
An iron abundance analysis yields ${\rm [Fe/H]}= 0.16\pm0.07$~dex. The object is thus
metal rich compared to the Sun. We also find substantial rotational broadening of the
spectral lines, the value of $v \sin i$ can be constrained in the range $4.5$ to
$7.5~{\rm km \, s}^{-1}$. These parameters are consistent with (or very close to) those
determined by L\`ebre et al.~(1999) and do~Nascimento~et al.~(2000). 
It thus appears that HD~137510 was a metal rich late F-type star which has now started 
its evolution away from the main sequence. Interestingly, Smith, Cunha, \& Lazzaro~(2001) 
have noted that HD~137510 possibly underwent chemically fractioned accretion and represents a good 
candidate for having a close-in giant planetary companion.
 
\section{Radial velocity data and Keplerian orbital solution}

We carried out observations of HD~137510 as part of the ongoing Doppler surveys
for extrasolar planets using the Harlan J. Smith 2.7~m telescope at McDonald Observatory (McD) in West Texas 
and the 2.0~m Alfred Jensch Telescope at Th\"uringer Landessternwarte Tautenburg (TLS) in Germany. 
In September 1998 a first spectrum 
of HD~137510 was taken at McD, but RV monitoring was discontinued for almost three years. 
In May 2001 observations of HD~137510 were started at TLS, while at the same time monitoring
was resumed at McD. For both programs we use a temperature stabilized molecular iodine vapor (I$_2$) cell
as the velocity metric. To obtain the high precision differential RV measurements we employ specialized 
data modeling techniques, which include the detailed reconstruction of the instrumental profile of the 
respective spectrometers at the time of observation 
(Valenti, Butler, \& Marcy~1995; Butler et al.~1996; Endl, K\"urster, \& Els~2000).

\begin{figure} 
\plotone{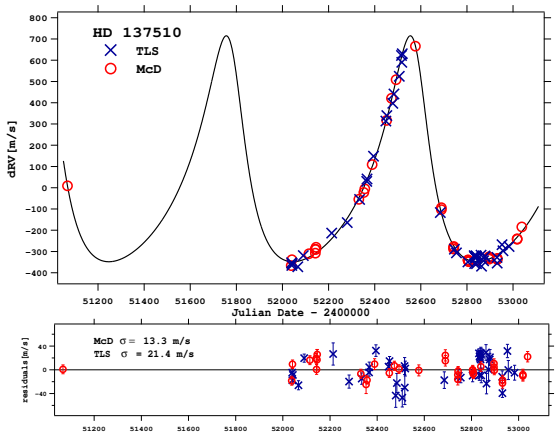}
\caption{RV data for HD~137510 (G0IV), from McDonald Observatory (McD) and Tautenburg
Observatory (TLS). The best-fit combined Keplerian orbital solution (see table~\ref{orbittab}
for the orbital parameters) is plotted as solid line. The lower panel shows the residuals
after subtraction of this orbit.
        }
\label{orbit_time}
\end{figure}

Fig.~\ref{orbit_time} displays our RV data as a function of time and with the best-fit 
Keplerian orbital solution superimposed. 
We employ the program $Gaussfit$ (McArthur, Jefferys, McCartney~1994), which uses a robust estimation method 
to find orbital solutions for each data set (McD and TLS) independently, as well as a combined solution, 
where the arbitrary velocity zero points ($\gamma$ velocities) of both data sets are included 
as fit parameters. 
Table~\ref{orbittab} summarizes the parameters we determined for the orbit. The resulting
mass function of this binary yields a minimum mass for the companion of $26\pm1.4~{\rm M}_{\rm Jup}$.        
Table~\ref{vels} lists our Doppler measurements for HD~137510. 

The residual scatter around this orbital fit is unusually high, McD: $\sigma=13.3~{\rm m \, s}^{-1}$ and TLS: 
$\sigma=21.4~{\rm m \, s}^{-1}$, roughly a factor of $2$ worse 
than the routine RV precision
for each telescope system. It also exceeds the average internal measurement errors of both data sets 
(McD: $8.9~{\rm m \, s}^{-1}$, TLS: $12.0~{\rm m \, s}^{-1}$).   
Stellar activity can introduce additional noise in RV data at that level of measurement precision (e.g.
Saar \& Donahue~1997). For HD~137510 we can rule out this possibility, based on the general low Ca II H\&K 
emission of this star. The McD spectra, for which we determined the RVs, also include the Ca II H\&K lines,
two well known activity indicators. We find a McD S-index ${\rm S}_{\rm McD}=0.157 \pm 0.011$ 
(see section 2.3 in Paulson et al.~2002 for 
a detailed description of how we obtain the S-index for the Ca II K line core emission). 
This is a comparable emission level as the well known 
inactive star $\tau$~Ceti (${\rm S}_{\rm McD}=0.166 \pm 0.008$) and shows that 
HD~137510 is not an active star.    
Furthermore, the star does not show up as an X-ray source in the ROSAT all-sky survey results
(H\"unsch, Schmitt, \& Voges~1998). 
One possible explanation for the higher scatter is the substantial rotational broadening of the 
spectral lines in the case of HD~137510 ($v \sin i =4.5 - 7.5~{\rm km \, s}^{-1}$). The
larger scatter for the TLS data likely reflects the differences both in telescope aperture 
(TLS: $2.0$~m ; McD: $2.7$~m) as well as in the relative seeing conditions at these two sites.   
Another possible explanation is the presence of a second low-level Keplerian signal. We searched
in the RV residuals for periodic signals using the Lomb-Scargle periodogram (Lomb~1976, Scargle~1982).
While the two separate periodograms of the individual sets of residuals do not reveal a significant signal, 
we do find a strong signal with a period of $227$ days in the combined residuals. This warrants 
further intensive monitoring of HD~137510 to confirm a possible planetary companion at a 
separation of $\approx0.8$~AU (which would make this system very interesting from a dynamical standpoint).   

\section{Discussion}

Due to the general sparseness of brown dwarf RV companions we have to look for additional
clues in order to rule out a stellar mass companion producing the detected Keplerian signal:

Simply based on the probabilty of a random distribution of the inclination angle $i$, we already have 
a $90\%$ probability that $i$ is larger than $25^{\circ}$, which translates into a true mass of
$<61~{\rm M}_{\rm Jup}$, still below the hydrogen burning limit.   

We measure a $v \sin i$ of $4.5$ to $7.5~{\rm km \, s}^{-1}$ for HD~137510, 
which is a typical value for a G0IV star. The mean $v \sin i$ value of the F8 - G2 subgiants 
in the de Medeiros \& Mayor (1999) catalogue is $6.5~{\rm km \, s}^{-1}$ (excluding one fast 
rotating F8IV star the mean value drops to $5.7~{\rm km \, s}^{-1}$). 
This shows that the value for HD~137510 is not atypically low, which 
means that we probably do not have a near pole-on view on the primary star.
Assuming that the orbital plane is 
roughly perpendicular to the rotation axis of the star we, again, do not seem to 
face a low $i$ scenario in this case.

McAlister et al.~(1987) obtained a negative result for this star in their speckle survey for
duplicity among Bright Star Catalogue stars using the CFHT. 
Granted, a positive detection would have been a difficult one, close to the diffraction limit of the telescope 
($\approx 0.04$ arcsec).   

HD~137510 has no flag in the multiple/binary annex of the $Hipparcos$
catalogue, showing that no suspicious astrometric variation was discovered.
According to Lindegren et al.~(1997) a stochastic solution is
chosen for objects where no acceptable single or double star solution can be found and additional noise,
typically ranging from $3$ to $30$ mas, needs to be added until an acceptable fit is found. If we regard 
the $3$ mas as the lower limit of an astrometric pertubation (angular size of the semimajor axis) which 
would have shown up in the analysis of the $Hipparcos$ data of a bright target like HD~137510, we can 
exclude inclination angles of $i < 16^{\circ}$ and a true
mass of the companion of $>94~{\rm M}_{\rm Jup}$, using the constraint of Pourbaix \& Jorissen (2000).

\section{Conclusion}
We have presented observational evidence for the substellar nature of the 
companion to HD~137510. {\it The question, why this star constitutes one of the very rare oases  
in the brown dwarf desert, remains unanswered.} 
However, due to the well established parameters of the primary and the orbit, 
this new system might help to understand binary formation processes 
for these mass-ratios at small orbital separations.   

\acknowledgements
We are grateful to the McDonald Observatory Time-Allocation-Committee for generous allocation of 
observing time. This material is based upon work supported by
the National Aeronautics and Space Administration under Grant NAG5-9227 issued
through the Office of Space Science, and by National Science Foundation
Grant AST-9808980. We thank our referee Adam Burgasser for his helpful comments.

\begin{deluxetable}{lrrr}
\tabletypesize{\scriptsize}
\tablecaption{
Orbital parameters for the companion to HD~137510. Solutions are given for both 
data sets separately, as well as the combined solution.}
\tablewidth{0pt}
\tablehead{
\colhead{} & \colhead{McD} & \colhead{TLS} & \colhead{McD+TLS}
\label{orbittab} }
\startdata
Period [days]			& $799.0\pm 1.4$ 	&$804.5\pm23.2$		& $798.2 \pm 1.4$ \\
T$_{\rm periastron}$ [JD]	& $2452583.2\pm3.2$ 	&$2452572.3\pm9.5$	& $2452582.01 \pm 2.6$\\
K[${\rm m\,s}^{-1}$] 		& $531.7\pm6.9$ 	&$508.9\pm22.5$		& $531.6 \pm 5.3$\\
$e$ 				& $0.40\pm0.008$ 	&$0.397\pm0.03$		& $0.402\pm0.008$\\
$\omega$ 			& $30.9\pm1.4$ 		&$28.7\pm3.1$		& $30.8 \pm  1.2$\\
$m\sin i~[{\rm M}_{\rm Jup}]$ 	& $26\pm1.4$ 		&$25\pm1.4$ 		& $26\pm1.4$\\
$a$ [AU] 			& $1.85\pm0.05$ 	&$1.86\pm0.05$		& $1.85\pm0.05$ \\ 
$rms\,[{\rm m\,s}^{-1}]$ 	& $13.1$ 		& $21.0$		& McD:$13.3$ TLS:$21.4$ \\
\enddata
\end{deluxetable}

\begin{deluxetable}{rrrr}
\tabletypesize{\scriptsize}
\tablecaption{
Differential radial velocities for HD~137510 (note that both
data sets have arbitrary and thus different velocity zero-points).}
\tablewidth{0pt}
\tablehead{
\colhead{JD [2400000+]} & \colhead{dRV [${\rm m\,s}^{-1}$]} & 
\colhead{error [${\rm m\,s}^{-1}$] } & \colhead{Survey}
\label{vels} }
\startdata
51067.6406 & 165.7  & 7.5 & McD \\
52037.7836 & -209.7 & 7.5 & McD \\
52040.8490 & -182.1 & 7.1 & McD \\
52114.7432 & -154.0 & 7.5 & McD \\
52141.6609 & -133.2 & 8.2 & McD \\
52142.6541 & -151.1 & 8.3 & McD \\
52144.7022 & -133.4 & 17.4 & McD \\
52145.6828 & -123.6 & 8.3 & McD \\
52331.9521 & 101.9 & 7.7 & McD \\
52353.9910 & 134.5 & 15.2 & McD \\
52357.9625 & 151.6 & 6.6 & McD \\
52388.9414 & 265.5 & 9.1 & McD \\
52451.7614 & 473.5 & 7.4 & McD \\
52472.7426 & 577.6 & 7.6 & McD \\
52492.6720 & 665.2 & 7.4 & McD \\
52576.5355 & 822.8 & 9.3 & McD \\
52689.0031 & 62.4  & 9.4 & McD \\
52689.0124 & 53.4  & 8.9 & McD \\
52741.8983 & -124.1 & 11.9 & McD \\
52741.9064 & -131.8 & 9.4 & McD \\
52743.9543 & -119.9 & 7.7 & McD \\
52804.8642 & -184.4 & 7.3 & McD \\
52804.8749 & -183.0 & 7.5 & McD \\
52807.6845 & -191.4 & 7.7 & McD \\
52807.6914 & -190.1 & 7.5 & McD \\
52839.7200 & -186.1 & 8.3 & McD \\
52895.5874 & -174.6 & 8.5 & McD \\
52895.6010 & -168.5 & 8.0 & McD \\
52896.6072 & -179.5 & 7.4 & McD \\
52931.5547 & -176.0 & 8.2 & McD \\
52931.5686 & -180.9 & 8.4 & McD \\
53018.9949 & -83.5  & 9.2 & McD \\
53018.0065 & -85.7  & 8.8 & McD \\
53038.9840 & -27.5  & 8.7 & McD \\ 
 & & & \\
52039.5781 &     -220.5  &       7.1 & TLS \\
52040.5039 &     -233.4   &       6.4  & TLS \\
52042.4922 &     -224.5  &       6.4  & TLS \\ 
52066.4219 &     -239.6  &       7.6  & TLS \\ 
52090.4375 &     -186.8  &       6.8  & TLS \\ 
52213.2188 &      -81.1   &      18.7  & TLS \\
52280.6875 &      -31.6   &      11.0 & TLS \\ 
52334.5703 &       77.4  &       9.5 & TLS \\ 
52365.5430 &      162.0   &       6.3 & TLS \\
52366.5469 &      173.5  &       8.6 & TLS \\
52394.5391 &      280.6  &       9.7 & TLS \\
52448.4219 &      445.9  &       5.7 & TLS \\
52452.3984 &      471.8  &       7.7  & TLS \\ 
52478.4805 &      529.5   &      19.8 & TLS \\
52483.4062 &      573.5   &      16.1  & TLS \\ 
52506.3320 &      656.7  &      15.7 & TLS \\
52516.3125 &      754.5  &      14.3 & TLS \\
52517.2969 &      722.5  &      28.6 & TLS \\
52519.3086 &      762.8  &      12.5 & TLS \\
52684.6445 &       15.5  &      14.0 & TLS \\
52745.5664 &     -156.9  &      11.0 & TLS \\ 
52754.5312 &     -174.1  &       8.5 & TLS \\
52804.4180 &     -217.1  &      10.3 & TLS \\
52831.3984 &     -187.3  &       7.6  & TLS \\
52833.3984 &     -224.4  &       8.5 & TLS \\
52834.4102 &     -193.4  &      10.5 & TLS \\
52835.3984 &     -192.3  &       8.4 & TLS \\
52836.4766 &     -195.8  &      10.2 & TLS \\
52838.4766 &     -198.3  &      14.6 & TLS \\
52839.4141 &     -187.7  &       9.9 & TLS \\
52840.4023 &     -190.5  &      10.8 & TLS \\
52841.3516 &     -225.3  &      12.3 & TLS \\
52861.3750 &     -185.1  &      13.2 & TLS \\
52863.3516 &     -237.1  &      17.1 & TLS \\ 
52872.3398 &     -211.5  &      18.7 & TLS \\
52877.3672 &     -193.1  &      10.3 & TLS \\ 
52878.3594 &     -189.0  &      13.0 & TLS \\
52930.2266 &     -195.6  &       9.8 & TLS \\
52931.2227 &     -222.3  &       7.2 & TLS \\
52952.1992 &     -135.6  &      11.2 & TLS \\
52956.1992 &     -164.3  &      16.1 & TLS \\  
52982.7031 &     -144.2  &      12.5 & TLS \\
\enddata
\end{deluxetable}

\end{document}